\documentclass[pra,aps,twocolumn,floats,floatfix,nofootinbib,superscriptaddress]{revtex4-1}

\usepackage{amsmath}
\usepackage{amssymb}
\usepackage{amsfonts}
\usepackage{graphicx}
\usepackage{bm}

\newcommand{\threed}{\text{3d}}
\newcommand{\kin}{\text{kin}}
\newcommand{\pot}{\text{pot}}
\newcommand{\up}{\uparrow}
\newcommand{\down}{\downarrow}
\newcommand{\vek}[1]{\bm{\mathrm{#1}}}
\newcommand{\vv}{\vek{v}}
\newcommand{\rv}{\vek{r}}
\newcommand{\pv}{\vek{p}}
\newcommand{\energy}{\varepsilon}
\newcommand{\nablav}{\vek{\nabla}}
\newcommand{\Eq}[1]{Eq.~(\ref{#1})}
\newcommand{\Eqs}[1]{Eqs.~(\ref{#1})}
\newcommand{\Fig}[1]{Fig.~\ref{#1}}
\newcommand{\Ref}[1]{Ref.~\cite{#1}}
\newcommand{\Refs}[1]{Refs.~\cite{#1}}
\newcommand{\Sec}[1]{Sec.~\ref{#1}}
\newcommand{\eq}{\text{eq}}
\newcommand{\ho}{\text{ho}}
\newcommand{\rep}{\text{rep}}

\begin{document}
\title{Shock waves in colliding Fermi gases at finite temperature}

\author{S. Chiacchiera}
\affiliation{Science and Technology Facilities Council
(STFC/UKRI), Daresbury Laboratory, Keckwick Lane, Daresbury, Warrington WA4
  4AD, United Kingdom}

\author{D. Davesne}
\affiliation{Universit\'e Lyon 1, Universit{\'e} de Lyon, CNRS-IN2P3,
  Institut de Physique Nucl{\'e}aire de Lyon, 43 Bd. du 11 Novembre
  1918, F-69622 Villeurbanne cedex, France}

\author{M. Urban}
\email{urban@ipno.in2p3.fr}
\affiliation{Institut de Physique Nucl{\'e}aire, CNRS-IN2P3,
  Universit{\'e} Paris-Sud, and Universit{\'e} Paris-Saclay, 91406
  Orsay Cedex, France}
                     
\begin{abstract} 
  We study the formation and the dynamics of a shock wave originating
  from the collision between two ultracold clouds of strongly
  interacting fermions as observed at a lower temperature in an
  experiment by Joseph {\it et al.} \cite{Joseph2011}. We use the
  Boltzmann equation within the test-particle method to describe the
  evolution of the system in the normal phase. We also show a direct
  comparison with the hydrodynamic approach and insist on the
  necessity of including a shear viscosity and a thermal conductivity
  term in the equations to prevent unphysical behavior from taking
  place.
\end{abstract}

\maketitle

\section{Introduction}
Shock waves in a fluid are very strong variations (almost
discontinuities) in density, pressure or temperature, that travel
through the system. In the case of classical gases, they have been
studied for a long time, both within the framework of the partial
differential equations of hydrodynamics (Navier-Stokes equations)
\cite{Becker1922} and within kinetic theory (Boltzmann equation)
\cite{Mott-Smith1951}. Since the width of the shock front is typically
just a few times the mean free path, hydrodynamics is actually not
well suited for a quantitative description of the shock front
\cite{Mott-Smith1951}.

Joseph et al. \cite{Joseph2011} observed shock waves in an ultra-cold
strongly interacting Fermi gas of trapped $^6$Li atoms. In this
experiment, a cigar-shaped atom cloud was initially split into two in
the axial direction; then, the barrier was removed and the two halves
were let to collide in the center of the common harmonic potential. It
was observed that the evolution towards the new equilibrium was quite
violent in the first stages: the density profile developed a sharp
peak at the center of the trap that later expanded in a ``box-like''
shape with clear edges. In some sense, this collision of two
degenerate Fermi gas clouds resembles the collision of two heavy ions
where the possibility of shock wave formation was suggested
\cite{Scheid1973,Danielewicz1979}.

Joseph et al. \cite{Joseph2011} showed that the behaviour observed in
their cold-atom experiment could be nicely reproduced in the framework
of quasi one-dimensional (1d) viscous hydrodynamics at zero
temperature, and they used this to estimate the viscosity by fitting
the width of the shock front. An alternative description of the
experiment, that involves a dispersive rather than a dissipative
mechanism to control the steepening, was presented in
\Refs{Salasnich2011,Ancilotto2012,Wen2015}. In these works, the authors derive
the superfluid hydrodynamic equations from an extended Thomas-Fermi
density functional, and the viscosity term is not needed. At present,
also due to spatial resolution issues, both descriptions are
compatible with the available data and the discussion about the
(almost) zero temperature case is not settled yet. Experimental
methods to distinguish between the two mechanisms, dissipative and
dispersive, were proposed in \Ref{Lowman2013}. A more microscopic
interpretation of the experiment was proposed in \Ref{Bulgac2012} using
the so-called time-dependent superfluid local-density approximation
(TDSLDA). There, the violent dynamics leads to rapid phase
oscillations of the order parameter which do not die out even at late
times, which is probably a consequence of the lack of true dissipation
in the TDSLDA framework.

In the present work, we use a numerical solution of the Boltzmann
equation to simulate shock waves in colliding Fermi gases. In this
way, we do not rely on the validity of (viscous) hydrodynamics and we
can study the relaxation towards thermal equilibrium at late
times. However, we cannot simulate the experiment that is done deeply
in the superfluid regime with an initial temperature close to zero,
but we consider a normal-fluid (but still degenerate) gas at finite
temperature.

In \Sec{sec:method}, the framework of the Boltzmann equation and its
numerical solution are briefly summarized. In \Sec{sec:collision}, we
show numerical results for the collision of two clouds similar to the
experiment of \Ref{Joseph2011}. In \Sec{sec:anisotropy}, we
concentrate on the anisotropy of the momentum distribution in the
shock front. Nevertheless, as shown in \Sec{sec:hydro}, it turns out
that a hydrodynamical model is able to reproduce most of the Boltzmann
results. Finally, conclusions are drawn in \Sec{sec:conclusions}.

Unless otherwise stated, we use units with $\hbar=k_B=1$, where
$\hbar$ and $k_B$ are the reduced Planck constant and the Boltzmann
constant, respectively.

\section{Boltzmann equation and its numerical solution}
\label{sec:method}
Let us briefly summarize our approach, for details see
\Refs{Lepers2010,Pantel2015}. We consider a three-dimensional (3d) gas
of fermionic atoms of mass $m$, having two ``spin'' states $\up$ and
$\down$ (in reality these may be two hyperfine states) with equal
populations and interacting {\it via} a short-range interaction
characterized by the $s$-wave scattering length $a$. The system is
trapped in an external potential $V$.

To describe the dynamics, we start from the semiclassical Boltzmann
equation for the distribution function $f(\rv,\pv,t) =
f_\up(\rv,\pv,t) = f_\down(\rv,\pv,t)$ depending on coordinate $\rv$,
momentum $\pv$, and time $t$ \cite{Landau10}
\begin{equation}\label{eq:boltz}
\frac{\partial f}{\partial t}+\frac{\pv}{m}\cdot\nablav_{\rv}f
  +\vek{F}\cdot\nablav_{\pv}f =-I[f]\,.
\end{equation}
On the left-hand side, $\vek{F} = -\nablav V$ is the force due to the
external potential. In a more detailed quantitative study, $V$ should
also include a mean-field shift as discussed in
\Refs{Chiacchiera2009,Pantel2015}. We made some preliminary tests and
found that this attractive energy shift, which gets weaker with
increasing temperature, has the effect to slightly reduce the cloud
size and the propagation speed of the shock. However, it does not
qualitatively change the results, but it makes it more difficult to
check, e.g., the thermalization in the radial direction, which will be
discussed in \Sec{sec:anisotropy}. Therefore, we neglect the
mean-field shift in this work.

On the right-hand side of \Eq{eq:boltz}, $I[f]$ is the collision
integral that takes into account the effect of elastic two-body
collisions,
\begin{multline}
I[f] = \int \frac{d^3p_1}{(2\pi)^3} \int d\Omega \frac{d\sigma}{d\Omega}
      \frac{|\pv - \pv_1|}{m} \\
      \times \left[ f f_1 (1-f') (1-f'_1) 
        -  f' f'_1 (1 - f) (1 - f_1) \right].
  \label{eq:colli}
\end{multline}
In the first term in square brackets, $\pv$ and $\pv_1$ are the
initial and $\pv'$ and $\pv'_1$ the final momenta, while in the second
term the roles are exchanged. For the distribution functions we adopt
the short-hand notation $f = f(\rv,\pv,t)$, $f_1 = f(\rv,\pv_1,t)$,
$f' = f(\rv,\pv',t)$, and $f'_1 = f(\rv,\pv'_1,t)$. The Pauli blocking
of final states is expressed by the factors $1-f'$ etc. As a
consequence of momentum and energy conservation, $\pv'$ and $\pv'_1$
are uniquely determined by $\pv$ and $\pv_1$ and two angles denoted
$\Omega$. In the present case, the differential cross section
$d\sigma/d\Omega = 1/[1/a^2+(\pv-\pv_1)^2/4]$ is independent of the
scattering angle $\Omega$ in the center-of-mass frame. In
\Refs{Riedl2008,Chiacchiera2009,Pantel2015}, in-medium corrections to
the cross section were considered. These corrections tend to increase
the collision rate, i.e., to reduce the mean free path. In the context
of shock waves, one expects that this would mainly affect the width of
the shock front. However, like the mean field, these corrections get
weaker with increasing temperature and their effect is limited to a
small region of the cloud where the density is highest. Since the use
of the in-medium cross section in the Boltzmann calculation
considerably increases the computation time and its effect turned out
to be relatively weak in our previous studies of collective modes
\cite{Chiacchiera2011,Pantel2015}, we neglect it here.

The static (i.e. equilibrium) solution of \Eqs{eq:boltz} and
(\ref{eq:colli}) is the Fermi function
\begin{equation}
f_\eq = \frac{1}{e^{(p^2/2m+V-\mu)/T}+1}\,,
\end{equation}
where $T$ is the temperature and $\mu$ the chemical potential fixed by
the total number of atoms $N=2/(2\pi)^3\, \int d^3r\,d^3p\,f$.

To solve \Eq{eq:boltz} numerically, we employ the test-particle
method. This amounts to replacing the continuous distribution function
$f$ by a sum of a large number of $\delta$ functions (``test
particles''),
\begin{equation}
\label{eq:testparticles}
f \propto \sum_i \delta[\rv-\rv_i(t)]\delta[\pv-\pv_i(t)]\,,
\end{equation}
which are initially distributed randomly according to the equilibrium
distribution $f_\eq$ and then follow their classical trajectories
$\rv_i(t), \pv_i(t)$ in the potential $V$ except when they collide. A
collision takes place whenever the distance of two trajectories $i$
and $j$ in their point of closest approach is less than the distance
given by the cross section. The new momenta $\pv'_i$ and $\pv'_j$ are
then determined by rotating the initial momenta $\pv_i$ and $\pv_j$ in
the center-of-mass frame by a random angle $\Omega$. In order to
satisfy the Pauli principle, the collision of two test particles $i$
and $j$ is only accepted with a probability
$[1-\tilde{f}(\rv_i,\pv'_i)][1-\tilde{f}(\rv_j,\pv'_j)]$, where
$\tilde{f}$ is the convolution of \Eq{eq:testparticles} with Gaussians
of appropriate widths in $\rv$ and $\pv$ spaces in order to get a
continuous result.

Let us mention that this numerical method was successful in describing
experimental results for the anisotropic expansion
\cite{Cao2011,Elliott2014} and collective oscillations
\cite{Riedl2008} at a quantitative level, throughout the transition
from collisional hydrodynamics at temperatures slightly above the
superfluid-normal transition to the collisionless regime at high
temperature \cite{Pantel2015}. A very similar method \cite{Goulko2012}
was also used to describe the different regimes from bouncing to
transmission observed in the collision of two fully polarized clouds
\cite{Goulko2011}.

\section{Simulation of colliding clouds}
\label{sec:collision}
To study the shock wave, we simulate the experimental procedure of
\Ref{Joseph2011}. Initially, the system is in equilibrium in a potential
$V = V_{\ho}+V_{\rep}$ that is the sum of an elongated harmonic
potential
\begin{equation}
V_{\ho}(\rv) = \frac{m}{2}[\omega_\perp^2 (x^2+y^2)+\omega_z^2 z^2]\,,
\end{equation}
and a repulsive barrier along the axial direction,
\begin{equation}
V_{\rep}(\rv) = V_0 e^{-z^2/\sigma_z^2}\,,
\end{equation}
splitting the cloud into two. The parameters of the simulation are
inspired from those of the experiment \cite{Joseph2011} and are
summarized in Table \ref{tab:param}, where (we exceptionally keep
factors of $\hbar$ and $k_B$ for clarity) $\bar\omega =
(\omega_\perp^2\omega_z)^{1/3}$ is the average trap frequency, $l_{ho}
= (\hbar/m\bar{\omega})^{1/2}$ the corresponding oscillator length,
$E_F = (3N)^{1/3} \hbar\bar{\omega}$ the Fermi energy, $T_F = E_F/k_B$
the Fermi temperature, and $k_F = (2mE_F)^{1/2}/\hbar$ the Fermi
momentum.
\begin{table}
\caption{\label{tab:param} Parameters of the simulation (following
  \Ref{Joseph2011}).}
\begin{ruledtabular}
\begin{tabular}{lll}
$m$ ($^6$Li)       & (u)            & $6.015$               \\
$N$                &                & $2\times 10^5$         \\
$\omega_\perp/2\pi$ & (Hz)           & $437$                 \\
$\omega_z/2\pi$    & (Hz)           & $27.7$                \\
$\bar{\omega}$     & (ms$^{-1}$)     & $1.095$              \\
$l_{\ho}$           & ($\mu$m)       & $3.106$              \\
$V_0/k_B$              & ($\mu$K)       & $12.7$               \\
$\sigma_z$         & ($\mu$m)       & $21.2$               \\
$T_F$              & ($\mu$K)       & $0.706$              \\
$k_F$              & ($\mu$m$^{-1}$) & $4.181$              \\
$T/T_F$ (initial)  &                & $0.3$                \\
$(k_Fa)^{-1}$       &                & $-0.1$
\end{tabular}
\end{ruledtabular}
\end{table}

While our simulation is fully three-dimensional (3d), the trap
geometry with $\omega_z \ll \omega_\perp$ results in an almost
one-dimensional (1d) behaviour in the sense that the equilibration in
the transverse direction is fast compared to the timescale relevant
for the motion in $z$ direction. It is therefore useful to introduce
quantities that are integrated over $\rv_\perp = (x,y)$, such as the 1d
density,
\begin{equation}
  n(z) = \int d^2r_\perp n_{\threed}(\rv)
  = 2\int d^2r_\perp \int \frac{d^3p}{(2\pi)^3} f(\rv,\pv)\,,
\end{equation}
(the factor of 2 is the spin degeneracy) and the 1d velocity,
\begin{equation}
v(z)
  = \frac{2}{n(z)} \int d^2r_\perp
  \int \frac{d^3p}{(2\pi)^3} \frac{p_z}{m} f(\rv,\pv)
 \equiv \frac{\langle p_z\rangle}{m}\,.
\label{eq:v1d}
\end{equation}

In order to make the two clouds collide, the barrier $V_{\rep}$ is
removed. Because of the harmonic potential $V_{\ho}$, the two clouds
are then accelerated towards each other and after some time they
collide in the center of the trap ($z=0$).

\begin{figure}
\includegraphics[width=7.8cm]{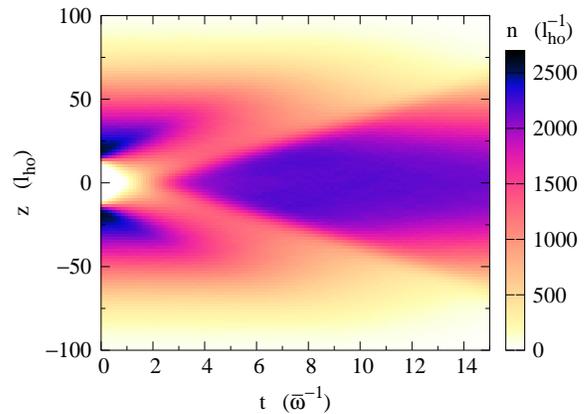}
\caption{\label{fig:n1d} 1d density profile $n(z,t)$ as function
  of position $z$ (vertical axis) and time $t$ (horizontal axis) after
  the removal of the barrier.}
\end{figure}

In \Fig{fig:n1d}
we show the 1d density profile $n(z,t)$ as a function of $z$ and
$t$. After $t\sim \bar{\omega}^{-1} \sim 1$ ms, the two clouds start
to touch each other, and at $t\sim 2$ ms the density starts to develop
a peak at $z=0$, as in the experiment \cite{Joseph2011}. At later
times, the peak expands and the density inside becomes flat. There is
a clear separation between the central region with high density and
the outer part with lower density. As a function of time, this step
moves outwards and was identified with a shock front.

\begin{figure*}
\includegraphics[width=15.6cm]{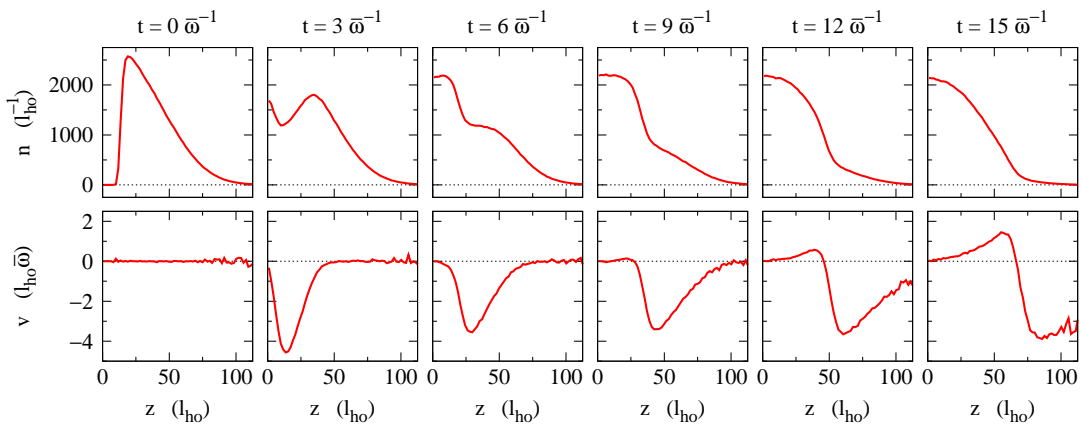}
\caption{\label{fig:nv1d} Snapshots of the density profile $n(z)$
  (top) and velocity profile $v(z)$ (bottom) at different times $t$
  after the removal of the barrier. Only positive $z$ values are
  shown, since $n(-z) = n(z)$ and $v(-z) = -v(z)$ for symmetry
  reasons.}
\end{figure*}

The box-like shape of the high-density region can be better seen in
the upper panels of \Fig{fig:nv1d}
which show snapshots of the density profile, especially at $t =
6\,\bar{\omega}^{-1}$ and $t = 9\,\bar{\omega}^{-1}$.  In the lower
panels of \Fig{fig:nv1d}, we show the velocity $v$. Although the
central region with high density and low velocity is clearly separated
from the outer part with lower density which continues to fall inwards
($v < 0$), $n$ and $v$ are of course not really discontinuous, but the
shock front has a finite width. Let us mention that the
numerical noise of average quantities per particle in the low-density
region corresponds to statistical fluctuations due to the test-particle method
and that the smooth aspect observed otherwise has been obtained by
averaging over several runs with different microscopic
initializations. More specifically, the Boltzmann results shown in
this work correspond to averages over 48 runs, each involving $5\times
10^4$ test particles.

In \Ref{Joseph2011}, this violent two-cloud collision with shock wave
formation was studied within a 1d superfluid hydrodynamic (i.e.,
zero-temperature) model. In order to obtain the finite width of the
shock front, the Euler equation was extended to include a viscous
force, and the viscosity was determined by fitting the experimental
density profile. However, the viscous force should in principle be
accompanied by dissipation, i.e., heating, in order to conserve the
total energy. In our case, since we start already from a finite
temperature, it is clear that for the hydrodynamic description one has
to solve three coupled equations, corresponding to the conservation of
particle number, momentum, and energy. It turns out that not only the
viscosity but also the heat conductivity is crucial if one wants to
reproduce the Boltzmann results, see \Sec{sec:hydro}.

\section{Momentum anisotropy in the shock front}
\label{sec:anisotropy}
It was already pointed out long time ago \cite{Mott-Smith1951}
that hydrodynamics, even with viscosity and heat conductivity, is not
applicable in the shock front. To see this in the present case, we
compare the following moments of the distribution function
\begin{gather}
  \energy_{\kin,x}(z) = \frac{1}{2m}\langle p_x^2\rangle\,,\\
  \energy_{\pot,x}(z) = \frac{m\omega_\perp^2}{2}\langle x^2\rangle\,,\\
  \energy_{\kin,z}(z) = \frac{1}{2m}(\langle p_z^2\rangle -
                        \langle p_z\rangle^2)\,,
\end{gather}
where the averages $\langle\cdots\rangle$ are defined as in
\Eq{eq:v1d}. Because of axial symmetry, we have of course
$\energy_{\kin,y} = \energy_{\kin,x}$ and $\energy_{\pot,y} =
\energy_{\pot,x}$ (with $\energy_{\kin,y}$ and $\energy_{\pot,y}$
defined analogously to $\energy_{\kin,x}$ and
$\energy_{\pot,x}$). In the 1d hydrodynamic model, one would assume a
local equilibrium in the sense that $\energy_{\kin,x} =
\energy_{\pot,x} = \energy_{\kin,z} = \energy/5$ with $\energy$
the internal energy per particle in the local fluid rest frame. In a
classical gas, $\energy_{\kin,x}$ and $\energy_{\kin,z}$ can be identified with
``anisotropic temperatures'' $T_\perp$ and $T_z$ in transverse and
longitudinal directions, respectively.

\begin{figure*}
  \includegraphics[width=15.6cm]{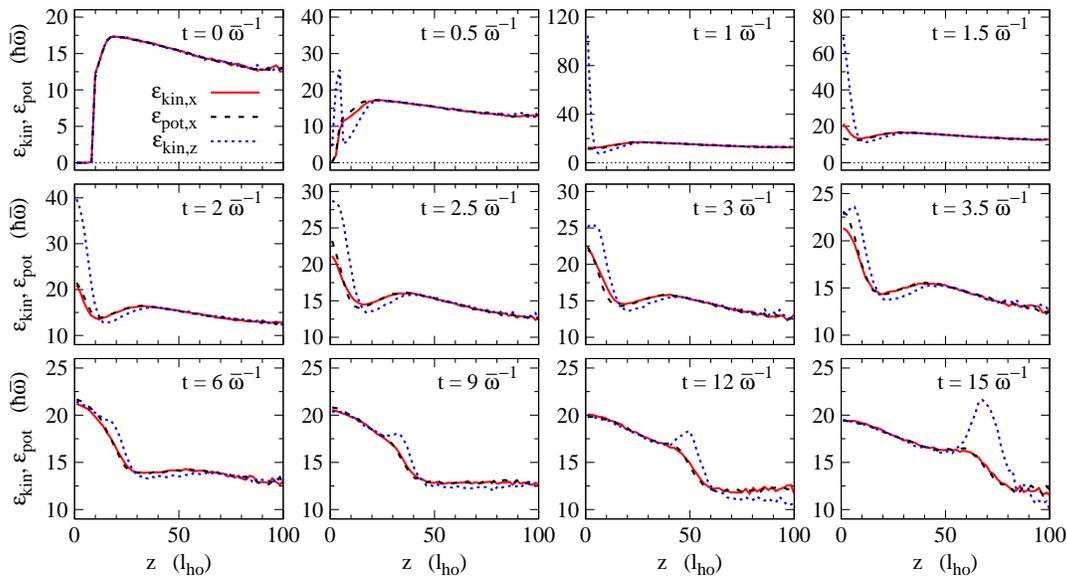}
  \caption{\label{fig:ei} Moments $\energy_{\kin,x}(z)$ (red solid
    lines), $\energy_{\pot,x}(z)$ (black dashes) and
    $\energy_{\kin,z}(z)$ (blue dots) as functions of $z$ (calculated
    in 200 bins) at different times $t$ after the removal of the barrier.}
\end{figure*}

In \Fig{fig:ei}
we show $\energy_{\kin,x}(z)$, $\energy_{\pot,x}(z)$, and
$\energy_{\kin,z}(z)$ for different times after the removal of the
barrier. One can clearly see that the equipartition in the transverse
direction, $\energy_{\kin,x} = \energy_{\pot,x}$, is very well
fulfilled, except for a fast oscillation of $\energy_{\kin,x}$ against
$\energy_{\pot,x}$ in the beginning of the collision. This is,
however, not the case for the longitudinal direction. One clearly sees
that $\energy_{\kin,z} > \energy_{\kin,x}$ in the region of the shock
front, even at large times when the shock front reaches the
low-density region. Far away from the shock, the gas is in local
equilibrium and one finds $\energy_{\kin,x} = \energy_{\kin,z}$. Far
outside, the gas becomes classical with $\energy_{\kin,i} = T/2$ (the
initial temperature is $T = 25.3\, \hbar\bar{\omega}$). In the center,
the relation between $\energy_{\kin,i}$ and the temperature is more
complicated because the gas is degenerate and $\energy_{\kin,i}$ is
dominated by Fermi motion.

In fact, the result $\energy_{\kin,z} > \energy_{\kin,x}$ can be intuitively
understood: particles entering the zone of the shock from the dilute
region need to undergo a couple of collisions before their
distribution corresponds to the equilibrium one of the dense
region. Because of the different $\langle p_z\rangle$ in the two
regions, the variance of $p_z$ in the transition region is obviously
larger than the variance of $p_x$.

Within this picture, the ansatz for a bimodal momentum distribution
used in \Ref{Mott-Smith1951} looks very natural: there, the momentum
distribution in the shock front is written as a superposition of
equilibrium distributions corresponding to the mean velocities and
temperatures on both sides of the shock. Recent high-precision
numerical calculations confirm this picture, see
e.g. \Ref{Malkov2015}.

However, when we analysed our momentum distribution in the shock front
we did not find any bimodality. This may have two reasons: first, our
gas is not classical but degenerate, which leads to a substantial
broadening of the momentum distribution, and second, the shock we are
studying is not very violent. Therefore, one could maybe describe the
anisotropy of the momentum distribution in the shock front within the
framework of the so-called ``anisotropic fluid dynamics''
\cite{Bluhm2015}.

\section{1d hydrodynamic model}
\label{sec:hydro}
In the original experimental paper \cite{Joseph2011} the shape of the
shock front was used to extract an effective viscosity
parameter. Therefore, it is interesting to study how far one can get
with viscous hydrodynamics in spite of the anisotropy of the momentum
distribution in the shock front discussed in the preceding section. In
our case, we assume that the system is in the normal-fluid phase at
finite temperature, which makes the hydrodynamic description somewhat
more complicated and requires to include also the heat conductivity.

Starting from the usual (3d) hydrodynamic equations
(cf. \Ref{Landau6}, Eqs. (15.5) and (49.2)), integrating over $x$ and
$y$, assuming that the system remains always in equilibrium in the
transverse direction [i.e., $T_{\threed}(\rv) = T(z)$ and
  $\mu_{\threed}(\rv) = \mu(z)-m\omega_\perp^2r_{\perp}^2/2m$] and
neglecting the contribution of the transverse components of the fluid
velocity $\vv_{\threed}$ to the energy, one obtains the following 1d
hydrodynamic equations:
\begin{gather}
  \dot{n} = -(n v)'\,,\label{hydro1}\\
  (n v)^{\!\mbox{$\cdot$}} = -\Big(n v^2+\frac{P}{m}\Big)'+n \frac{F}{m}
    + \frac{4}{3m}(\eta v')'\,,\label{hydro2}\\
  (n E)^{\!\mbox{$\cdot$}} = -(n v E + v P)' + n v F
    + \Big(\frac{4}{3}\eta v v'+\kappa T'\Big)'\,,\label{hydro3}
\end{gather}
describing the conservation of particle number, momentum, and
energy. In these equations, $E=\energy+\tfrac{1}{2}mv^2$ is the total
energy per particle, $P$ is the pressure integrated over $x$ and $y$,
$F=-m\omega_z^2 z$ is the $z$ component of the external force, and
$\kappa$ and $\eta$ are respectively the heat conductivity and shear
viscosity integrated over $x$ and $y$ (in the presence of a
non-vanishing bulk viscosity $\zeta$, one would have to replace
$\tfrac{4}{3}\eta$ by $\tfrac{4}{3}\eta+\zeta$). Derivatives are
denoted as $\dot{v} = \partial v/\partial t$ and $v' = \partial
v/\partial z$ etc.

To close this system of equations, it is necessary to express $P$,
$T$, $\eta$ and $\kappa$ in terms of $n$ and $\energy$. Since we
neglect the mean field, we consider the equation of state (EOS) of an
ideal gas, $P = (\gamma-1)n\energy$, with $\gamma = \tfrac{7}{5}$ as
in a gas of diatomic molecules because the internal energy $\energy$
includes five degrees of freedom ($\energy_{\kin,x}$,
$\energy_{\kin,y}$, $\energy_{\kin,z}$, $\energy_{\pot,x}$,
$\energy_{\pot,y}$). To find $T$, it is convenient to write $n$ and
$P$ as functions of $\mu$ and $T$ in the form $n =
A\,T^{5/2}F_{3/2}(\mu/T)$ and $n\energy = A\,T^{7/2}F_{5/2}(\mu/T)$,
with the abbreviation $A = 4\sqrt{2m}/(3\pi\omega_\perp^2)$ and the
integrals of the Fermi function $F_\alpha(x) = \int_0^\infty
\!dt\,t^\alpha/(e^{t-x}+1)$. The functions $F_\alpha$ and their
inverse functions $X_\alpha$ [i.e., $X_\alpha(F_\alpha(x)) = x$] can
be very efficiently approximated \cite{Antia1993} and then $T$ is
obtained by solving numerically the equation $X_{5/2}(n\energy/(A\,
T^{7/2})) = X_{3/2}(n/(A\,T^{5/2}))$. For $\eta$, we follow
\cite{Joseph2011} and assume that $\eta = \tilde{\eta} n$ with some
constant $\tilde{\eta}$, although this choice cannot easily be
justified.\footnote{Notice that one cannot compute $\eta$ by
  integrating $\eta_{\threed}$ from kinetic theory over $x$ and $y$:
  since in kinetic theory $\eta_{\threed}\propto T^{3/2}$ depends only
  on temperature and not on density, this integral would not even
  converge.} Similarly, we also assume that $\kappa = \tilde{\kappa}
n/m$. The dimensionless (for $\hbar=k_B=1$) proportionality constants
$\tilde{\eta}$ and $\tilde{\kappa}$ are fitted to give reasonable
agreement with the Boltzmann results. In the notation of
\Ref{Joseph2011}, our $\tilde{\eta}$ corresponds to
$\tfrac{3}{4}m\nu$. For completeness, we recall that in
\Ref{Joseph2011} a value of $\tilde{\eta}=7.5$ was obtained at
(almost) zero temperature.

Although \Eqs{hydro1}--(\ref{hydro3}) can formally be obtained by
integrating the 3d hydrodynamic equations over $x$ and $y$, the
equipartition of the energy among the five internal degrees of freedom
observed in the Boltzmann results seems to indicate that the range of
validity of the 1d equations is actually larger than that of the 3d
equations which must fail at large $x$ and $y$ where no collisions
take place.
\begin{figure*}
  \includegraphics[width=15.6cm]{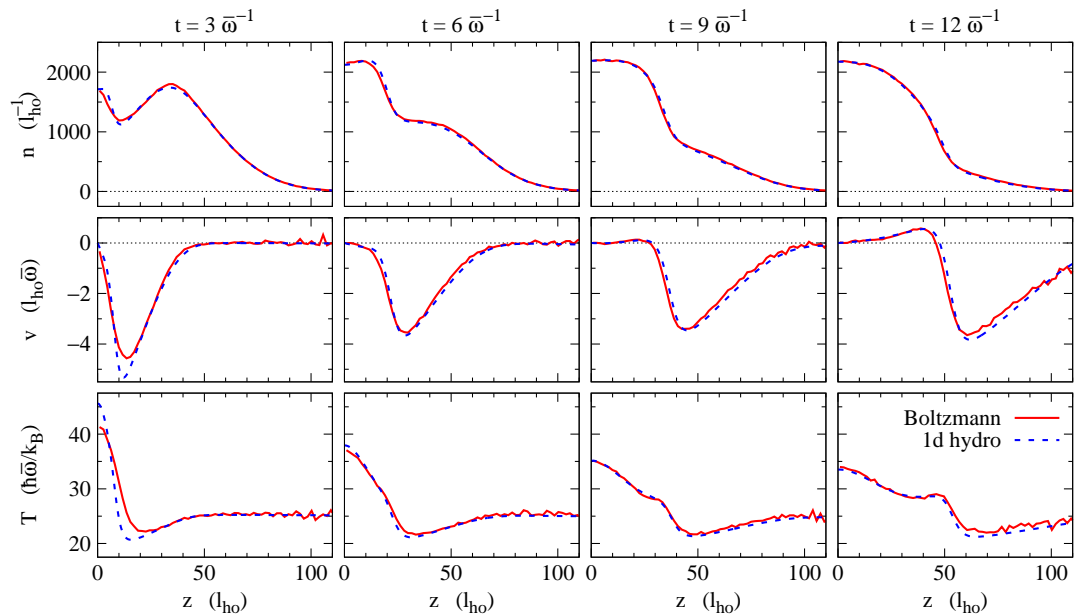}
\caption{\label{fig:hydro-Boltzmann} Comparison between 1d
  hydrodynamic (blue dashes) and Boltzmann (red solid lines) results
  for the density $n$, the velocity $v$, and the temperature $T$ in
  trap units (from top to bottom) at four different times $t = 3, 6,
  9$ and $12\,\bar{\omega}^{-1}$. The dimensionless viscosity and heat
  conductivity parameters are $\tilde{\eta} = 9$ and $\tilde{\kappa} =
  34$. The Boltzmann results for the temperature were obtained using
  the EOS of the ideal Fermi gas with harmonic radial confinement (see
  text) with the Boltzmann results for $n$ and $\energy$ as input.}
  \end{figure*}
\begin{figure}
  \includegraphics[width=7.8cm]{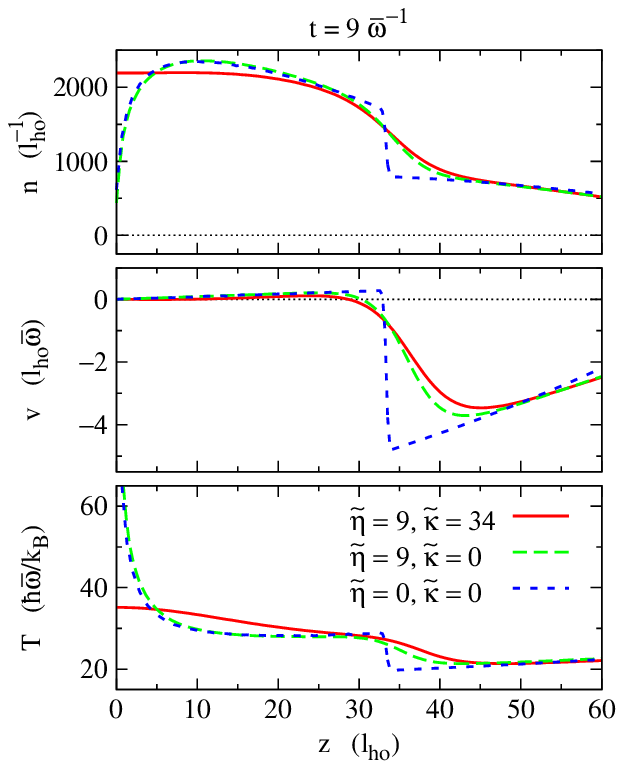}
\caption{\label{fig:eta-kappa} Results of 1d hydrodynamics for the
  density $n$, the velocity $v$, and the temperature $T$ in trap units
  (from top to bottom) at $t = 9\,\bar{\omega}^{-1}$. Red solid line:
  results obtained with $\tilde{\eta} = 9$ and $\tilde{\kappa} = 34$;
  green long dashes: $\tilde{\eta}=9$ but no heat conductivity; blue
  short dashes: no viscosity and no heat conductivity.}
\end{figure}

The numerical solution of \Eqs{hydro1}--(\ref{hydro3}) is
nontrivial. Because of the instabilities of standard methods in the
case of shock waves, dedicated methods have been developed. Here we
use the code ``hydro1d'' \cite{Zingale-hydro1d} based on Riemann
solvers \cite{Toro} which we have extended to include the external
force in axial direction due to the trap potential, the above
mentioned EOS of the Fermi gas with harmonic radial confinement, and
the viscosity $\eta$ and heat conductivity $\kappa$. The viscous and
heat conduction terms, like the external force, were implemented as
source terms added to the equations in conservative form (similar to
the implementation of gravity in the original hydro1d code).

Except at small times ($t\lesssim 5\,\bar{\omega}^{-1}$), an excellent
agreement between hydrodynamic and Boltzmann results can be achieved
with $\tilde{\eta} = 9$ and $\tilde{\kappa} = 34$ as can be seen in
\Fig{fig:hydro-Boltzmann},
showing various snapshots for the density $n$, velocity $v$ and
  temperature $T$ profiles.
When determining these values, we have assumed that the ratio
$\eta/\kappa$ is approximately given by the ratio of
$\eta_{\threed}/\kappa_{\threed} = \tfrac{4}{15}m$ obtained in kinetic
theory \cite{Bruun2007,Braby2010}.

The importance of the heat conductivity becomes clear from
\Fig{fig:eta-kappa}, where we focus on a single snapshot, and compare
different hydrodynamics results for the profiles of $n$, $v$ and $T$.
There, the red solid curves represent the full results with viscosity
and heat conduction, while the long green dashes were obtained without
heat conduction and the short blue dashes without heat conduction and
viscosity. While the viscosity alone, without heat conduction, is
sufficient to smoothen the discontinuity across the shock front, it
does not remove the unrealistic hole in the density profile at
$z=0$. This hole is related to the temperature peak at $z=0$ which is
created during the initial stage of the collision. The hole in $n$ and
the peak in $T$ lead to a pressure that is practically constant around
$z=0$, so that there is no acceleration of matter. Hence, without heat
conduction, the peak in $T$ remains there forever and so does the hole
in $n$. But notice that, even with $\tilde{\kappa}=34$ the temperature
peak melts more slowly within 1d hydrodynamics than within Boltzmann
(cf. \Fig{fig:hydro-Boltzmann}, $t = 3\,\bar{\omega}^{-1}$).

The sharp shock front in the case $\tilde{\eta}=\tilde{\kappa}=0$
allows us to extract quantitatively some information which is
difficult to obtain from the smooth shock front in the realistic
case. In particular, one can read off the speed $v_s$ of the shock
front from $v_s = (v_1 n_1-v_2 n_2)/(n_1-n_2)$ where $v_i$ and $n_i$
are the velocities and densities on both sides of the
discontinuity. Using this, one can define the Mach numbers
$\text{Ma}_i = |v_i-v_s|/c_i$ where $c_i$ is the speed of sound on
side $i$ of the discontinuity. The upstream Mach number (i.e., the one
on the right-hand side of the discontinuity if one
looks at $z>0$), which quantifies the strength of the shock, is
$\text{Ma} \approx 2.6$ at $t=2\,\bar{\omega}^{-1}$. It decreases
rapidly to $\text{Ma} \approx 1.6$ at $t=6\,\bar{\omega}^{-1}$. At
$t\gtrsim 12\,\bar{\omega}^{-1}$, when the shock reaches the
low-density region, it starts to increase again.

\section{Conclusions}
\label{sec:conclusions}


In this paper, we applied the Boltzmann equation within the
test-particle method to the description of the dynamics of a violent
collision between two ultracold clouds as described in
\Ref{Joseph2011}. Within our approach, it is not possible to simulate
exactly the experimental conditions since the initial temperature in
the experiment is such that the system is superfluid. We thus focused
on the dynamics of the normal phase at higher temperature. In that
respect we have described in detail the formation and propagation of
the shock front as observed in \Ref{Joseph2011}. A direct comparison
with a 1d hydrodynamic approach clearly indicates that the extraction
of a viscosity parameter directly from a fit of the density profile is
probably doubtful and that the inclusion of a thermal conductivity is
unavoidable if one wants to get closer to the Boltzmann simulation and
prevent from unphysical behavior.

In this paper, we were mainly interested in the comparison between
Boltzmann and hydrodynamic calculations. Therefore we consistently
used in the Boltzmann calculation the propagation in the trap
potential without mean field, and in the hydrodynamic calculation the
equation of state of the ideal Fermi gas. For a quantitative
comparison with recent experiments done at finite temperatures
\cite{Roof2018}, it will be necessary to go beyond these
approximations, i.e., to include the mean field in the Boltzmann
calculation and the equation of state of the unitary Fermi gas
\cite{Ku2012} in the hydrodynamic calculation. As already mentioned in
\Sec{sec:method}, also the in-medium modification of the cross section
in the collision term should be taken into account. We are currently
working on these extensions of the present study.

As a perspective for future work, the emergence of the effective 1d
hydrodynamic behavior needs to be better understood from the
microscopic point of view. In particular, the effective 1d viscosity
and heat conductivity cannot be obtained by integrating the
corresponding 3d quantities over $r_\perp$ and should be derived from
a transport theory. Since even in the shock front the momentum
anisotropy seems to be moderate, it may be also possible to achieve
this goal within the recently developed anisotropic fluid dynamics
\cite{Bluhm2015}. Furthermore, to address also the situation of the
gas being, at least partly, in the superfluid phase, a more
sophisticated theory would be needed that combines the hydrodynamic
description of the superfluid with a transport equation for the
normal-fluid component.

\section*{Acknowledgments} 
S.~C. thanks J.~Castagna and G.~Romanazzi for valuable discussions,
and IPN Orsay for kind hospitality. We thank the Laboratory for
Advanced Computing at the University of Coimbra (Portugal) for
providing CPU time in the Navigator cluster.


\end{document}